\documentclass[12pt]{iopart}

\usepackage{graphicx}
\usepackage[T1]{fontenc} 
\usepackage{subfigure} 
\usepackage{amsfonts}

\newcommand{\Eqref}[1]{equation~(\ref{#1})}

\begin{document}
\title{Impurity Entanglement in the $J-J_2-\delta$ Quantum Spin Chain}
\author{Andreas Deschner and Erik S. S{\o}rensen}
\address{  Department of Physics and Astronomy, McMaster University 1280 Main
Street West, Hamilton, Ontario L8S 4M1, Canada}
\ead{deschna@mcmaster.ca}
\date{\today}

\begin{abstract}
The contribution to the entanglement of an impurity attached to one end of a
$J-J_2-\delta$ quantum spin chain ($S=1/2$) is studied.  Two different measures
of the impurity contribution to the entanglement have been proposed: the
impurity-entanglement-entropy $S_{\mathrm{imp}}$ and the negativity ${\cal N}$.
The first, $S_{\mathrm{ imp}}$, is based on a subtractive procedure where the
entanglement-entropy in the absence of the impurity is subtracted from results
{\it with} the impurity present.  The other, ${\cal N}$, is the negativity of a
part of the system separated from the impurity and the impurity itself.  In
this paper we compare the two measures and discuss similarities and differences
between them.  In the $J-J_2-\delta$ model it is possible to perform very
precise variational calculations close to the Majumdar-Ghosh-point ($J_2 = J /
2$ and $\delta = 0$) where the system is gapped with a two-fold degenerate
dimerized ground-state. We describe in detail how such calculations are done
and how they can be used to calculate ${\cal N}$ as well as $S_{\mathrm{ imp}}$
for {\it any} impurity-coupling $J_K$.  We then study the complete cross-over
in the impurity entanglement as $J_K$ is varied between 0 and 1 close to the
Majumdar-Ghosh-point. In particular we study the impurity entanglement when a
staggered nearest-neighbour-interaction proportional to $\delta$ is introduced.
In this case, the two-fold degeneracy of the ground-state is lifted leading to
a very rapid reduction in the impurity entanglement as $\delta$ is increased.
\end{abstract}


\maketitle

\section{Introduction}
\label{sec:intro}
Entanglement in quantum spin chains has received considerable attention.
For gapless quantum spin chains detailed predictions~\cite{Cardy04} of
many aspects of the entanglement have been obtained from conformal field
theory (CFT).  For these (1+1) dimensional models a precise
understanding of the entanglement has proven invaluable for the
understanding of numerical techniques such as as the density-matrix
renormalization group~\cite{White92} (DMRG) as well as for the
development of new techniques.  The contribution  to the entanglement
arising from impurities has also attracted considerable interest
~\cite{Zhou06,laflorencie_boundary_2006,Cho06,Kopp07b,Hur07c,Hur07a,sorensen_impurity_2007,sorensen_quantum_2007,bayat_negativity_2010,sodano_kondo_2010,Eriksson_2011}.
Here the term impurities is used in the general sense and includes the
effects of boundary magnetic fields~\cite{Zhou06},
boundaries~\cite{laflorencie_boundary_2006}, qubits interacting with a
decohering environment~\cite{Cho06,Kopp07b,Hur07c,Hur07a} as well as
Kondo-like
impurities~\cite{sorensen_impurity_2007,sorensen_quantum_2007,bayat_negativity_2010,sodano_kondo_2010,Eriksson_2011}.
The presence of the impurity can lead to different conformally invariant
boundary conditions which can dramatically alter the entanglement.  For
a review see reference~\cite{affleck_entanglement_2009}.  Quantum spin
models have been used for studies of qubit teleportation and quantum
state
transfer~\cite{Bose03,Christandl04,Christandl05,Burgarth05a,Burgarth05b,Wojcik05,Plenio05,Venuti06,Venuti07},
where the change in entanglement arising from the impurities plays a
crucial role.

Usually the entanglement is defined in terms of the von Neumann
entanglement-entropy of a sub-system $A$ of size $l$ and reduced
density-matrix $\rho_A$ defined by~\cite{Neumann27,Wehrl78},
\begin{equation}
  S(l,L)\equiv -\Tr [\rho_A\ln \rho_A]\ ,
  \label{eq:vNS}
\end{equation}
where $L$ stands for the total system size.  It is known
\cite{laflorencie_boundary_2006} that the entanglement-entropy of
Heisenberg-spin-chains with open boundary conditions have a uniform
as well as an alternating part:
\begin{equation}
  S(l,L)=S_u(l,L) + (-1)^l S_a(l,L) \ , 
  \label{eq:unialtS}
\end{equation}
where the subscripts $u$ and $a$ stand for uniform and alternating
respectively.

\begin{figure}[b]
  \begin{center}
    \includegraphics[width=0.8\textwidth]{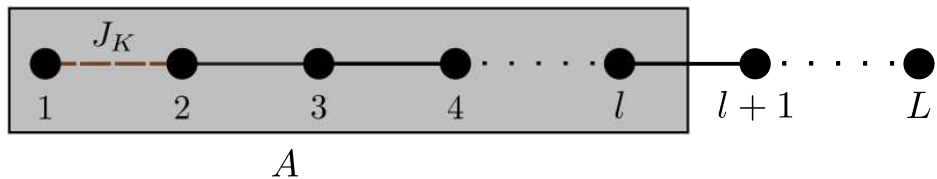}
  \end{center}
  \caption{An example of a spin chain with only
  nearest-neighbour-interactions and an impurity on the first site. The
  interactions of the impurity site are multiplied by a $J_K$.  The
  chain is separated into part $A$ (grey box), which includes the impurity
  and has the length $l$, and the rest of the chain. } 
  \label{fig:exentmea}
\end{figure}
Based on this observation, the impurity entanglement-entropy
$S_{\mathrm{imp}}(J_K, l, L)$ was introduced to quantify the
contribution an impurity has to the entanglement of a spin
chain~\cite{sorensen_impurity_2007,sorensen_quantum_2007}. To model an
impurity, the interactions on one site are scaled by a factor of $J_K$.
The impurity entanglement-entropy is given by the difference of the
uniform part of the entanglement-entropies of the chain with the
impurity present and the chain without the impurity. The sub-system $A$
in the definition of the entanglement-entropy in equation~\ref{eq:vNS}
is chosen to contain the first $l$ sites of the chain and thus contains
the impurity site, too (see figure~\ref{fig:exentmea}). Upon removing
the impurity site, $A$ as well as the whole chain are one element
shorter. The uniform entanglement-entropy in a system where the impurity
is absent is thus given by $S_{u}(1,l-1,L-1)$.
This leads to the following definition of $S_{\mathrm{imp}}$:
\begin{eqnarray}
  S_{\mathrm{imp}}(J_K,l,L)&\equiv& S_u({\rm with\ impurity})-S_u({\rm no\ impurity})\nonumber\\
  &\equiv& S_u(J_K,l,L)-S_{u}(1,l-1,L-1),\ l>1\ .
  \label{Simpdef}
\end{eqnarray}
This definition is similar in spirit to experimental procedures for
extracting impurity contributions to, for instance, susceptibilities. 
It is also possible to consider the alternating part of the impurity
entanglement but we shall not do that here.

However, it was later pointed out~\cite{bayat_negativity_2010,
sodano_kondo_2010} that a more consistent definition of the impurity
contribution to the entanglement can be obtained from the
negativity~\cite{vidal_computable_2002}. The negativity between a
sub-system $A$ and the rest of the system is defined by
\begin{equation}
  \mathcal{N} =\frac{ \sum | \lambda_i | - 1}{2}\ ,
  \label{Negdef}
\end{equation}
where the $\lambda_i$ are the eigenvalues of the partial transpose of
the density-matrix with respect to either $A$ or the rest of the system.
In contrast to the entanglement-entropy it can also be used to quantify
the entanglement if the system is \emph{not} in a pure state. This makes
it possible to use the negativity to directly quantify the entanglement
of the impurity-spin and another part of the system. 

It is an important question to what extent these two quantities yield
the same information about the impurity entanglement.  In this paper we
compare these two measures and show that they agree for many universal
features but are fundamentally different with $\mathcal{N}$ being the
more general measure of the impurity entanglement.

Entanglement and in particular impurity entanglement in gapped quantum
spin chains has received comparatively little attention. In part this is
due to the fact that for non-critical spin chains with a finite
correlation length $\xi$ one expects~\cite{Cardy04} for a partition into
2 semi-infinite chains:
\begin{equation}
S \sim \log{\xi}\ .
\end{equation}
For instance, for the $S=1$ Affleck-Kennedy-Lieb-Tasaki (AKLT)
chain~\cite{AKLT87,AKLT88} where the entanglement can be calculated
analytically~~\cite{Tribedi,Tribedi2,Hirano,Fath,Fan,Alipour,Scott} it
is known that $S(l,L)$ approaches a constant for large $l,L$ in an
exponential manner~\cite{Alipour,Scott} on a length scale equal to the
bulk correlation length $\xi=1/\ln(3)$. At the Majumdar-Ghosh-point
(MG-point) \cite{majumdar_antiferromagnetic_1970} of the $J-J_2$ spin
chain, where $J_2=J/2$, one finds that $S(l,L)$ is either 0 or $\ln(2)$
for any $l,L$.  However, the impurity contribution to the entanglement
{\it can still be long-range} in gapped spin chains. For the $J-J_2$
spin chain this was shown to be the case at the
MG-point~\cite{sorensen_impurity_2007,sorensen_quantum_2007}.  This is
due to the two-fold degeneracy of the singlet ground-state,
corresponding to the two different dimerization patterns, and the
associated presence of solitons separating regions with different
dimerization.  The model we study is the slightly more general
$J-J_2-\delta$ quantum spin chain defined
as:
\begin{eqnarray}
  H =&  J_K  \Big[(1+\delta)\vec{S}_0\cdot\vec{S}_{1} + J_2 \; \vec{S}_0\cdot\vec{S}_{2}\Big]  \nonumber \\
  &+\sum_{i=1}^{L-3}\Big[(1+(-\delta)^i)\vec{S}_i\cdot\vec{S}_{i+1} +J_2 \;
  \vec{S}_i\cdot\vec{S}_{i+2}\Big]  \nonumber \\
  &+ (1 + (-\delta)^{L-2}) \vec{S}_{L-2} \cdot \vec{S}_{L-1} \;  ,   
  \label{eq:hamiltonian}
\end{eqnarray} 
\begin{figure}[htb]
  \begin{center}
    \includegraphics[width=0.8\textwidth]{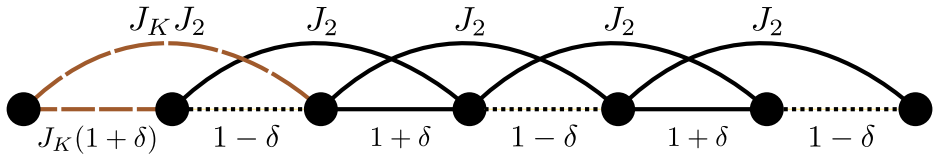}
  \end{center}
  \caption{ The $J-J_2-\delta$ chain with an impurity coupled with strength $J_K$.}
\end{figure}
where we implicitly have set the nearest-neighbour-coupling $J\equiv 1$.
The  coefficient $J_K$ describes the coupling of the impurity-spin
and $\delta$ is a staggering of the nearest-neighbour-coupling inducing
dimerization.  We focus exclusively on chains with open boundary
conditions and the impurity at one end of the chain.  With $\delta=0$
this model undergoes a transition from the gapless Heisenberg-phase to a
dimerized state at $J_2^c=0.241167$
\cite{haldane_spontaneous_1982,Eggert96}. In the Heisenberg-phase a
correspondence to the low-energy physics of a Kondo impurity
can be established~\cite{Eggert92,laflorencie_kondo_2008}. This correspondence becomes {\it exact} at $J_2^c$.
The mapping to the Kondo-problem requires one to use open boundary conditions and
we therefore do not consider periodic boundary conditions here.
However, entanglement in models similar to the one considered here but
with periodic boundary conditions have also been studied, see section 4 of Ref.~\cite{affleck_entanglement_2009}
for a review.

With periodic boundary conditions and an even number of sites, the
system at the MG-point has an exactly known two-fold degenerate dimerized
ground-state and a gapped spectrum
\cite{majumdar_antiferromagnetic_1970,shastry_excitation_1981}. (With open boundary conditions
and $L$ even the ground-state is unique with the same energy). The
lowest lying excitations of the MG-chain are pairs of unbound solitons
and upon introducing a finite dimerization the solitons become confined
\cite{sorensen_soliton_1998}. For a chain with an odd number of sites
solitons separate regions with the ground-state dimerization-pattern of
the chain with an even number of sites
\cite{shastry_excitation_1981,sorensen_soliton_1998}.  In the {\it
gapped } phase, in the proximity of the MG-point, it is possible to
perform very precise variational
calculations~\cite{shastry_excitation_1981,Caspers82,Caspers84} for the
$J-J_2$ model (corresponding to $\delta = 0 $ in this study) since
appropriate states spanning the variational sub-space can be identified
with relative ease, as the ground-state at the MG-point is known. Within
this variational framework it is possible to evaluate the
entanglement~\cite{sorensen_quantum_2007}. Here, we explain in detail
how similar calculations can be performed for the negativity and discuss
the appropriate sub-spaces used in these computations.  Along the so
called disorder-line, where $J_2=(1-\delta)/2$, one of the two
degenerate ground-states of the Majumdar-Ghosh-chain remains the {\it
exact} ground-state.  Close to the disorder-line, one therefore expects
the same variational methods to yield very precise results. 

In the past, these variational calculations have been performed more or
less by hand~\cite{shastry_excitation_1981,Caspers82,Caspers84}.  If the
variational problem is, however, formulated as a generalized
eigenvalue-problem, it is not difficult to automate the calculation, as
for any two states, $\varphi_i$ and $\varphi_j,$ with all but no, one or
two sites being in dimers,
$
\langle\varphi_i|\vec{S}_i \vec{S}_j|\varphi_j\rangle 
$ 
as well as the overlaps between different $\varphi$ can be computed in
an automated fashion.  One advantage of the automated computation of
these matrix-elements is that one avoids lengthy and error-prone
calculations.  Secondly, solving a generalized eigenvalue-problem can be
done using standard libraries.  This approach is not limited to the
specific Hamiltonian studied in this paper.  It can in fact be used to
perform variational calculations using dimerized states forming the
subspace in which the diagonalization is carried out for \emph{any}
Heisenberg-spin-Hamiltonian.

In addition to providing a more direct analytical insight into the
physics, the main advantages the variational approach has over other
methods such as the DMRG-technique are computational cost and
simplicity. This advantage does, however, not come for free. To identify
a good subspace to use for variational calculations, one always has to
rely on knowledge gained obtained by other means. This role is often
played by the DMRG-technique.

The paper is organized as follows:
In sections~\ref{sec:method}, \ref{sec:spaces}, \ref{sec:matele} and
\ref{sec:overlap} the variational approach is discussed along with the
different sub-spaces used in the calculations.  Section~\ref{sec:negativity}
describes the calculation of the negativity once the variational ground-state
is known and sample results for the negativity are compared to DMRG-data.  In
section~\ref{sec:negandsimp} the negativity and the
impurity-entanglement-entropy are compared.  Section~\ref{sec:negandjk}
presents our results for the negativity and $S_{\mathrm{imp}}$ for general
$J_K$ at the MG point.  In  section~\ref{sec:impentdeluneq0} we describe how
the impurity entanglement is affected by the presence of a non-zero explicit
dimerization, $\delta$.

\section{The Variational Calculation as a Generalized Eigenvalue-Problem}
\label{sec:method}
For completeness we review here how the variational problem can be formulated
as a generalized eigenvalue-problem.  Consider a set of states $\{\varphi_i\}$
to be used for the variational calculation. We then wish to minimize 
\begin{eqnarray}
  \langle H \rangle = \frac{( \varphi | H \varphi )}{(
  \varphi | \varphi )} \ ,
  \label{eq:var_definition}
\end{eqnarray}
where the trial-state $\varphi$ is given by:
\begin{eqnarray}
  \varphi = \sum c_j \varphi_j \ . 
  \label{eq:trialstate_definition}
\end{eqnarray}
The minimization is to be performed with respect to the $c_j$.  
In variational calculations we make use of the fact that:
\begin{eqnarray}
  \forall \varphi: \  E_0 &\le \frac{( \varphi |H \varphi)}{(\varphi|\varphi)} \ ,
\end{eqnarray}
where $E_0$ is the ground-state energy.  One strives to find a well chosen 
set of $\varphi_i$s as well as the state $\varphi_{min} \in \mathrm{span} \{
\varphi_i \}$ that satisfies
\begin{eqnarray}
  \frac{ ( \varphi |H \varphi) }{(\varphi|\varphi)} \Big{|}_{\varphi_{min}} &=
  \mathrm{min}_{ \mathrm{span} \{ \varphi_i \} } {  \frac{( \varphi |H
  \varphi)}{(\varphi|\varphi)} } \ .
  \label{eq:minimize1}
\end{eqnarray}
This state is the variational estimate for the ground-state.  Let us denote the
energy-eigenbasis with $\{ \mathfrak{e}_i\}$. The variational state $\varphi$
can evidently be written as $\varphi = \sum a_i \mathfrak{e}_i$.  Thus
\begin{eqnarray}
  \frac{( \varphi |H \varphi)}{(\varphi|\varphi)} &= \sum_{i,j} \bar{a}_i a_j
  \frac{(\mathfrak{e}_i|H\mathfrak{e}_j)}{\sum_{i}|a_i|^2} \nonumber \\
  &= \frac{\sum_{i}|a_i|^2 E_i }{\sum_{i}|a_i|^2} \nonumber \\
  & \ge  E_{v} \ , 
  \label{eq:vareig}
\end{eqnarray}
where $E_{v}$ is the energy of the eigenstate $\mathfrak{e}_{v}$ of non-zero
projection onto $\mathrm{span}\left\{ \varphi_i  \right\}$ that has the lowest
energy. With any orthonormal basis $\{ \mathfrak{b}_i \}$ of $\mathrm{span} \{
\varphi_i \}$, the variational guess for the ground-state $\varphi_{min}$ can
thus be found by diagonalizing $H$: 
\begin{itemize}
  \item 
    Minimization problem $\equiv$ Finding lowest eigenvalue of $\big(
    \mathfrak{b}_i| H \mathfrak{b}_j \big)$ 
\end{itemize}

If we do not have an orthonormal basis at our disposal, we have to do a little
more work.  Let us suppose that the $\varphi_i$ are linearly independent and
the $\mathfrak{e}_i$ are the orthonormal basis of energy-eigenstates.  For the
variational ground-state given by $\varphi_{min}= \sum_{i}a_i \varphi_i$ and
the variational estimate of the ground-state-energy $E_{v}$ we know from the
equations~(\ref{eq:vareig}) that
\begin{eqnarray}
  H \sum_{j}a_j \varphi_j &= E_{v} \sum_{j}a_j \varphi_j \ .
  \label{eq:minimize2}
\end{eqnarray}
We will now show that a vector $\mathfrak{a}=(a_1, \dots , a_n)$ is a solution
to the variational problem if and only if
\begin{equation}
  \forall  i: \; \; \sum_ja_j (\varphi_i| H \varphi_j) = E_{v} \sum_ja_j (\varphi_i| \varphi_j) \label{eq:geneigenval1} \ .
\end{equation}
With the definitions
\begin{eqnarray}
  \mathfrak{H}_{ij} = \big( \varphi_i\big|H \varphi_j\big) \phantom{und
  oder}\mathrm{and} \phantom{und oder} \mathfrak{B}_{ij} = \big( \varphi_i
  \big| \varphi_j \big)  \ , 
  \label{eq:hbdefi}
\end{eqnarray}
one can write \Eqref{eq:geneigenval1} as 
\begin{equation}
  \mathfrak{H}\mathfrak{a} = E_{v} \mathfrak{B} \mathfrak{a} \ ,
  \label{eq:geneigenval2}
\end{equation}
which defines a generalized eigenvalue-problem.
That \Eqref{eq:minimize2} implies \Eqref{eq:geneigenval1} can be seen by
taking the scalar-product of both sides in \Eqref{eq:minimize2} with
$\varphi_i$.

That equation~(\ref{eq:geneigenval1}) implies (\ref{eq:minimize2}) will now be
proved.  We do this by proving that for any basis (linearly independent set of
states that span the space) $\{\varphi_i\}$:
\begin{eqnarray}
  \forall  i: \; \; (\varphi_i| \mathfrak{q}) = ( \varphi_i | \mathfrak{r})\;\;\;
  \Rightarrow \;\;\; \mathfrak{q} = \mathfrak{r} \ ,
  \label{rel:linindequiv}
\end{eqnarray}
for any vectors $\mathfrak{q}$, $\mathfrak{r}$. With $\mathfrak{q} = H \sum_{j}a_j
\varphi_j$ and $\mathfrak{r} = E_{v} \sum_{j}a_j \varphi_j$ the relation
(\ref{rel:linindequiv}) becomes the relation equation~(\ref{eq:geneigenval1})
$\Rightarrow$ (\ref{eq:minimize2}).

Proof: 
\begin{eqnarray} 
  &  \forall  i: \; \; (\varphi_i| \mathfrak{q}) = ( \varphi_i |  \mathfrak{r}) \\ 
  \Rightarrow &\nonumber \\ 
  &  \forall  i: \; \; (\varphi_i|  \mathfrak{q}-\mathfrak{r}) = 0 \\ 
  \Rightarrow &\nonumber \\ 
  &  \forall  i: \; \;  \varphi_i \bot ( \mathfrak{q} - \mathfrak{r} ) \\
  \Rightarrow &\nonumber \\ 
  &   \phantom{\forall i: \;\;} \mathfrak{q} - \mathfrak{r} = 0  \phantom{ Beweis beended} \Box 
\end{eqnarray}

We have thus proven that the minimization-problem of \Eqref{eq:minimize1} can
be replaced by the generalized eigenvalue-problem of \Eqref{eq:geneigenval2},
if the states used to span the subspace in which the minimization takes place
are linearly independent.  

\section{The Hilbert Spaces Used in the Calculation}
\label{sec:spaces}
The implementation of the variational technique as a generalized
eigenvalue-problem allows us to just solve a linear algebra problem instead of
solving a set of non-linear equations which is necessary if one approaches the
minimization naively. The one requirement for this implementation to work is
for the states that span the Hilbert-space in which the minimization is done to
be linearly independent.  This flexibility makes it possible to use different
bases according to the specific needs of the problem. Here we are concerned
with the two cases of even and odd length chains.
\begin{enumerate}
  \item Chain with odd number of sites:\\
    The simplest ensemble of variational states for a chain with an odd number
    of sites are the so called single soliton
    states~\cite{shastry_excitation_1981,Caspers82,Caspers84,sorensen_quantum_2007}.
    These states can be generated by taking $L-1$ of the sites of the chain to
    be maximally dimerized while leaving the left over site (the soliton) to be
    in the $S_z=1/2$ state. Here and in the remainder of the paper we use $L$
    to denote the length of the chain. The number of dimers between the
    undimerized site  and the left end of the chain can be used to label the
    states.  In figure~\ref{fig:trailstatesodd1} we show a pictorial
    representation of the first three states. When drawing a dimer we use an
    arrow to denote the order in
    $(|\uparrow\downarrow\rangle-|\downarrow\uparrow\rangle)/\sqrt{2}$.  For an
    odd length chain with $L$ spins there exist $N=(L+1)/2$ such states.  Note
    that we here do not include states with the soliton in the $S_z=-1/2$
    state. 
    \begin{figure}[htb]
      \begin{center}
        \subfigure[$\varphi_{0}$]{\includegraphics[scale=0.75]{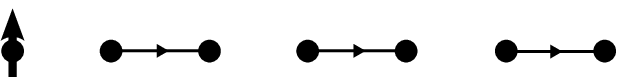}}\\
        \subfigure[$\varphi_{1}$]{\includegraphics[scale=0.75]{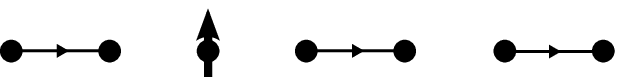}}\\
        \subfigure[$\varphi_{2}$]{\includegraphics[scale=0.75]{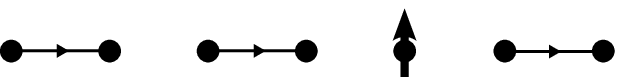}}
      \end{center}
      \caption{ The first three variational states used in the calculations
      with the single soliton space for a chain with an odd number of sites.
      Note the arrows on the dimers that indicate their direction.}
      \label{fig:trailstatesodd1}
    \end{figure}

  \item Chain with even number of sites:\\ 
    In studies of the entanglement of the impurity-site with the rest of
    a chain with an even number of sites, a suitable set of states can be
    defined by leaving the bulk $(L-2)$ sites of the chain dimerized while one
    of the spins is in a valence bond with the impurity.  Example-states
    are given in figure~\ref{fig:trailstateseven0}.
    \begin{figure}[htb] 
      \begin{center} 
         \subfigure[$\varphi_{0}$]{\includegraphics[scale=0.75]{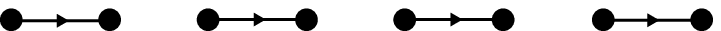}}\\
         \subfigure[$\varphi_{1}$]{\includegraphics[scale=0.75]{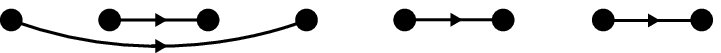}}\\ 
         \subfigure[$\varphi_{2}$]{\includegraphics[scale=0.75]{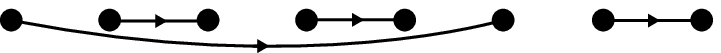}}
      \end{center} 
      \caption{ The first three variational states used in the calculation of
      the negativity for a chain with an even number of sites.}
      \label{fig:trailstateseven0} 
    \end{figure}
    These states are chosen in order to reflect the presence of an
    {\it impurity valence bond} (IVB)~\cite{sorensen_quantum_2007}, connecting
    the impurity site to the bulk of the chain.

\end{enumerate}
\section{The Matrix Elements of of the Hamiltonian $\mathfrak{H}$}
\label{sec:matele}
To set up the generalized eigenvalue-problem it is necessary to calculate the
matrices $\mathfrak{H}$ and $\mathfrak{B}$ which were defined in
\Eqref{eq:hbdefi}.  In this section we will explain how to compute
$\mathfrak{H}$.  In order to allow for the most general Hamiltonian possible we
write it as follows:
\begin{equation}
  H=\sum_{\{m,l\}}J_{ml} \vec{S}_m \vec{S}_l\ .
\end{equation}
Here, the summation $\{m,l\}$ is over 'bonds' of the Hamiltonian and the
coupling constants $J_{ml}$ can take any value.  It is convenient to write the
matrix-elements $\mathfrak{H}_{ij}$ in terms of the operators $h_{ml}
=\vec{S}_m \vec{S}_l-1/4$, because they have a simple action on the $\varphi$s.
This gives:
\begin{eqnarray}
  \big( \varphi_i\big|H \varphi_j\big) &= \sum_{\{m,l\}} J_{ml} \big(
  \varphi_i\big|\vec{S}_m \vec{S}_l \varphi_j\big)\nonumber\\ 
  &=
  \sum_{\{m,l\}} J_{ml}\big( \varphi_i\big|\big[h_{ml} +1 /
  4\big]\varphi_j\big) \ , 
\end{eqnarray}
The following description of this action is an extension of results presented
earlier \cite{beach_formal_2006} in the context of quantum Monte Carlo
simulations in the valence bond basis where it is necessary to take the
$h_{ml}$ to have the opposite sign of what we use here.  The action of $h_{ml}$
on states relevant to us is given by:
\begin{eqnarray}
  h_{ml}   \big[m;l\big] \big[k;n\big] & = -              \big[m;l\big] \big[k;n\big]  \label{eq:start_bondac} \\
  h_{mk}   \big[m;l\big] \big[k;n\big] & =    \frac{1}{2} \big[m;k\big] \big[n;l\big]  \label{eq:outside}      \\ 
  h_{ml}   \uparrow_m\big[l;k\big]     & =    \frac{1}{2} \big[m;l\big]\uparrow_k                              \\  
  h_{mk}   \big[m;l\big] \uparrow_k    & = -  \frac{1}{2} \Big[ \uparrow_m \big[l;k\big]+ \big[m;l\big]\uparrow_k \Big] \ , 
  \label{eq:end_bondac}
\end{eqnarray}
where
\begin{eqnarray*}
  \big[m;k\big]  &:= \frac{1}{\sqrt{2}} \big( \uparrow_m \downarrow_k - \uparrow_k \downarrow_m \big) 
\end{eqnarray*}
are the spins at $m$ and $k$ in a singlet-state.

\begin{figure}[h]
  \begin{center}
    \subfigure[]{\hspace{1.9cm} \includegraphics[scale=0.75]{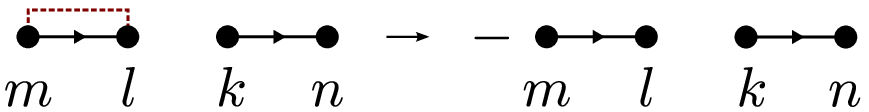}}
    \subfigure[]{\hspace{1.9cm} \includegraphics[scale=0.75]{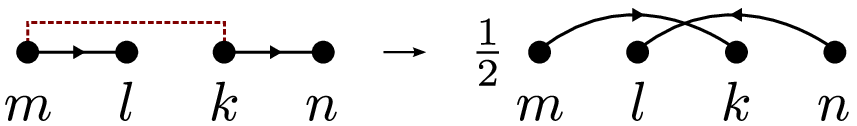}}
    \subfigure[]{\hspace{1.9cm} \includegraphics[scale=0.75]{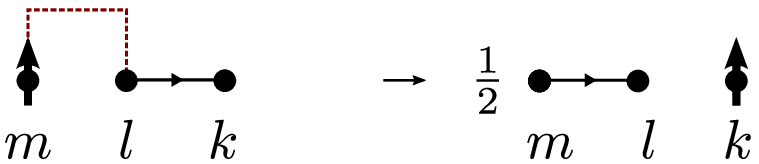}}
    \subfigure[]{\hspace{1.9cm} \includegraphics[scale=0.75]{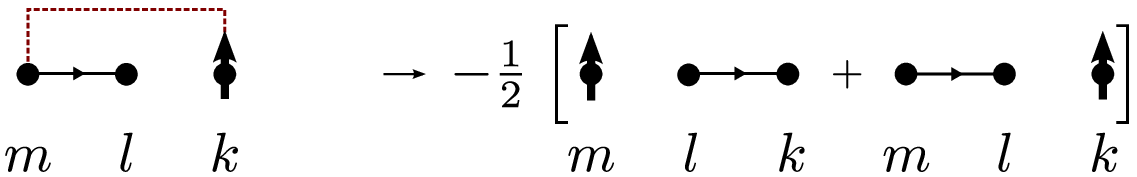}}
  \end{center}
  \caption{The action of $h_{ml}$ on the variational states. The dotted
  line signifies the sites acted upon by $h_{ml}$. Arrows fix the phase
  of singlets between sites of the lattice. The convention is explained
  in the main text.}
  \label{fig:bondac}
\end{figure}

In figure~\ref{fig:bondac} a pictorial representation of the
equations~(\ref{eq:start_bondac})-(\ref{eq:end_bondac}) is shown. To fix the
phase of the states we include in the calculation, we again represent
$\big[m;k\big]$ by an arrow pointing from $m$ to $k$.
For the calculations we present here these are the only rules needed.
However, when considering the action of the Hamiltonian it is sometimes
convenient to apply the simple rules
\begin{eqnarray}
  \big[m;k\big] & =-\big[k;m\big] \\
  \big[k;l\big]\uparrow_m & = \big[k;m\big]\uparrow_l+\uparrow_k\big[m;l\big] \ ,
  \label{eq:rules}
\end{eqnarray}
in order to reduce the action of the $h_{ml}$ to one of the above.

We want to stress that even though the single soliton states can be
orthogonalized with relative ease~\cite{uhrig_unified_1999} this is not
generally the case and even after orthogonalization the resulting eigenvalue
problem is still non-trivial for a general set of $J_{lm}$.  Whether or not it
is possible to proceed with analytical calculations often depends on specific
choices for the coupling constants in the Hamiltonian. Here, we can solve the
variational problem for any values of the couplings.

As can be seen in equation~(\ref{eq:outside}) and figure~\ref{fig:bondac}(b),
the application of $h_{ml}$ to a given state $\varphi_i$ may result in a state
that is not element of $\mathrm{span} \{ \varphi_i\}$.  Therefore, in order to
evaluate 
\begin{eqnarray*}
  \big( \varphi_i\big|H \varphi_j\big) &= \sum_{\{m,l\}} J_{ml} \big(
  \varphi_i\big|h_{ml} \varphi_j\big)
\end{eqnarray*}
we need to compute the overlap of $h_{ml} \varphi_j$ with all the variational
states.  This slows down our calculations since it is not sufficient to build a
table of all the overlaps of the $\varphi_i$. We now turn to a discussion of
how the overlaps are calculated.

\section{The Matrix Elements of the Overlap-Matrix $\mathfrak{B}$}
\label{sec:overlap}
As is explained in \cite{sutherland_systems_1988, beach_formal_2006} the
magnitude of the overlaps of valence-bond-states depends solely on the length
of loops created by overlaying the two states. If one chooses $l$ to be the
length of such a loop then the overlap is of magnitude $2^{-l/2 +1 }$.  To get
the sign of the overlap we, in addition to overlaying them,  reverse all arrows
on one of the states.  Then we follow the loop and count how often one has to
go against the arrow.  If the resulting number is odd, the sign of the overlap
is  negative.  If the resulting number is even, the sign of the overlap is
positive. Because the total number of bonds in the loop is even this leads to a
well defined sign of the overlap.
\begin{figure}[h]
  \begin{center}
    \subfigure[]{\includegraphics[scale=0.95]{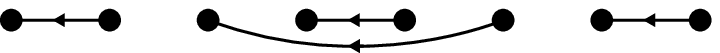}}\\
    \subfigure[]{\includegraphics[scale=0.95]{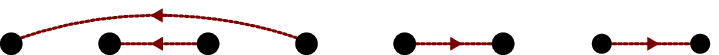}}\\
    \subfigure[]{\includegraphics[scale=0.95]{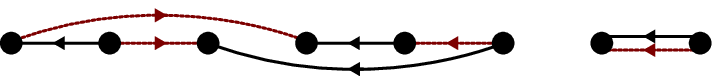}}
  \end{center}
  \caption{ The two  states (a) and (b) together with the corresponding
  loop-structure created by reversing the arrows of one state and
  overlaying the two states (c). Both loops contribute a minus-sign. The
  overlap is given by $1 / 4$.}
  \label{fig:transgraph}
\end{figure}
To get the overlaps of two states one only has to follow this prescription for
every loop and multiply contributions from different loops.
Figure~\ref{fig:transgraph}(c) shows the loop-structure for the states shown in
figure~\ref{fig:transgraph}(a) and figure~\ref{fig:transgraph}(b). The left
loop is six bonds long and going around it one has to go against the direction
of the arrow three times. The right loop is 2 bonds long and one has to go
against the direction of the arrow once.  The overlap is therefore given by:
\begin{eqnarray}
  (- 2^{-2})(-2^0) = \frac{1}{4} \ . \nonumber
  \label{}
\end{eqnarray}
In the case of unpaired spins being part of the states the rules have to be
modified slightly.  If there is one unpaired spin in both states, upon
overlaying the two states, there will in addition to loops be a string that
goes from one unpaired spin to the other. As is the case for the loops, for the
string the magnitude of the contribution to the overlap depends only on the
length.  One finds the contribution to be: $2^{-l/2}$.  The contribution to the
sign of the overlap can be determined in exactly the same way as in the case of
complete loops.

\section{The Calculation of the Reduced Density-Matrix and the Negativity}
\label{sec:negativity}
\begin{figure}[htb]
  \begin{center}
    \includegraphics[width=0.4\textwidth]{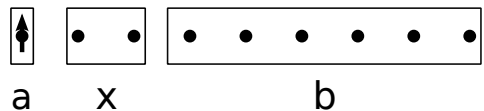}
  \end{center}
  \caption{The chain separated
  into three parts: the impurity-spin (a), the part whose entanglement with the
  impurity-spin is to be quantified (b) and the part that separates the two (x).}
  \label{fig:negpl}
\end{figure}

For the calculation of the negativity describing the entanglement between the
impurity-spin and a distant part of the chain, it is necessary to separate the
chain into three parts~\cite{vidal_computable_2002,bayat_negativity_2010}: the
impurity-spin, the part whose entanglement with the impurity-spin is to be
quantified and the part that separates the two (the parts will in the remainder
of the paper also be referred to as region a, b and x respectively. The
negativity is invariant under interchange of b and x.  For clarification see
figure~\ref{fig:negpl}).  
To calculate the negativity between the impurity-spin and a part of the chain
that is separated from the impurity-spin by a different part, we have to
compute the reduced density-matrix that results from tracing out the separating
part of the chain. We will now explain how this is done within our variational
framework. 

Let the result of the generalized eigenvalue problem be the variational
ground-state $\varphi_g$ given in the variational basis: $\varphi_g = \sum a_i
\varphi_i$.  To take trace of the density-matrix over the region x in
figure~\ref{fig:negpl}, we express the density-matrix in a product-basis formed
by states of region x and states of the rest of the chain. This requires us to
represent the variational ground-state $\varphi_g$ in this product-basis. In
doing so we follow the approach of reference~\cite{sorensen_quantum_2007},
where the entanglement-entropy was calculated from a variational ground-state.
For a chain with an odd number of spins these bases will in the following be
introduced. We start with the case that the region~x contains an even number of
sites followed by the case that the region~x contains an odd number of sites.
For a chain with an even total number of sites an analogous calculation was
performed.
\subsection{An even number of sites to be traced out}
\begin{enumerate}
  \item For the impurity-spin we choose the $S_z$-basis and denote the two
    states by $f_1=\, \uparrow $ and $f_2=\, \downarrow$ .
  \item For the region x we denote the basis by ${g_i}$. 
    We take the region to contain at least two sites.
    The states are chosen to be:
    \begin{eqnarray}
      g_1=& \sum_{n=1}^{R} a_n \  \underbrace{\uparrow\ \overbrace{\ \relbar
      \phantom{\uparrow} \dots \phantom{\uparrow} \relbar \ }^{n-1\ \mathrm{dimers}}\
      \uparrow \ \relbar \ \dots \ \relbar}_{2R\ \mathrm{ sites}} \  , \nonumber \\
      g_2=& \sum_{n=1}^{R} a_n  \ \underbrace{\uparrow\ \overbrace{\ \relbar
      \phantom{\uparrow} \dots \phantom{\uparrow} \relbar \ }^{n-1 \ \mathrm{dimers}} \
      \downarrow \ \relbar \ \dots \ \relbar}_{2R\ \mathrm{ sites}} \ , \nonumber \\
      g_3=& \underbrace{ \uparrow \ \relbar \ \dots \ \relbar \uparrow}_{2R\ \mathrm{sites}}\ , \ \ 
      g_4= \underbrace{ \downarrow \ \relbar \ \dots \ \relbar \downarrow}_{2R\ \mathrm{sites}} \ , \nonumber \\
      g_5=& \underbrace{ \downarrow \ \relbar \ \dots \ \relbar \uparrow}_{2R\ \mathrm{sites}} \ , \ \
      g_6= \underbrace{ \uparrow \ \relbar \ \dots \ \relbar \downarrow}_{2R\ \mathrm{sites}}\ , \nonumber \\
      g_7=& \underbrace{  \relbar \ \dots \ \relbar }_{2R\ \mathrm{sites}} \ , \nonumber
      \label{eq:xstatesoddchain}
    \end{eqnarray}
    where the symbol '$\relbar$' is used for valence-bonds between
    neighbouring sites and $R$ is the highest possible number of dimers
    that can could be formed in region~x.

  \item For region~b we define the states in the following way:
    \begin{eqnarray}
      \tilde{h}_1=& \underbrace{  \relbar \ \dots \ \relbar }_{L-2R - 1
      \ \mathrm{sites}}\nonumber\\
      \tilde{h}_2=& \sum_{n=R+1}^{N-1} a_n \  \underbrace{\uparrow\ \overbrace{\ \relbar
      \phantom{\uparrow} \dots \phantom{\uparrow} \relbar \ }^{n-R-1\ \mathrm{dimers}}\
      \uparrow \ \relbar \ \dots \ \relbar}_{L-2R-1\ \mathrm{ sites}}
      \nonumber\\
      \tilde{h}_3=& \sum_{n=R+1}^{N-1} a_n \  \underbrace{\downarrow\ \overbrace{\ \relbar
      \phantom{\uparrow} \dots \phantom{\uparrow} \relbar \ }^{n-R-1\ \mathrm{dimers}}\
      \uparrow \ \relbar \ \dots \ \relbar}_{L-2R-1\ \mathrm{ sites}}
      \nonumber
      \label{eq:bstatesoddchain}
    \end{eqnarray} 
    In order to express the density-matrix of the impurity-spin and the
    region~b in an orthonormal basis it is necessary to orthonormalize the
    basis $\{\tilde{h}_i\}$ as it is not orthogonal the way it is defined
    above: 
    \begin{eqnarray}
      (\tilde{h}_1|\tilde{h}_3)=\sum_{n=R+1}^{N-1} a_n \ (-2)^{-n+R+1} \left( \frac{-1}{\sqrt{2}} \right) \ .
    \end{eqnarray}
    We do this following the usual Gram-Schmidt procedure.  Our
    orthonormalized version is given by:
    \begin{eqnarray}
      h_1   &= \frac{\tilde{h}_1}{\sqrt{(\tilde{h}_1|\tilde{h}_1)}}
      \nonumber\\
      h_2   &= \frac{\tilde{h}_2}{\sqrt{(\tilde{h}_2|\tilde{h}_2)}}
      \nonumber\\
      h_3   &= \frac{ \tilde{h}_3 - ( \tilde{h}_3|h_1 ) h_1 }
      {\sqrt{(\tilde{h}_3|\tilde{h}_3) - |(\tilde{h}_3|h_1)|^2} }  \ .
    \end{eqnarray}

\end{enumerate}
For the product-space of the impurity-spin and region~b we use the basis-states:
\begin{eqnarray}
  k_1&=f_1 h_1,\ k_2 = f_1 h_2 ,\ k_3 = f_1 h_3,\ \nonumber \\
  k_4&=f_2 h_1,\ k_5 = f_2 h_2 ,\ k_6 = f_2 h_3 \nonumber \ . 
\end{eqnarray}
With these definitions, the variational ground-state can be written as:
\begin{eqnarray}
  \varphi_g = \sum a_n \varphi_n = \sum_{i,j} C_{ij} k_i g_j \ ,
\end{eqnarray}
where the matrix $C$ is given by:
\begin{eqnarray}
  C = 
  \left(
  \begin{array}{ccccccc}
    0 & \frac{\|\tilde{h}_1\|}{\sqrt{2}} & 0 & 0 & 0 & \frac{(\tilde{h}_3|h_1)}{2} & a_0 \|\tilde{h}_1\|\\
    0 & 0 & 0 &-\frac{\|\tilde{h}_2\|}{\sqrt{2}} & 0 & 0 & 0 \\
    0 & 0 & 0 & 0 & \frac{ \kappa }{2} & 0 & 0 \\
    - \frac{\|\tilde{h}_1\|}{\sqrt{2}} & 0 & -\frac{(\tilde{h}_3|h_1)}{2}  & 0 & 0 & 0 & 0 \\
    0 & 0 & 0 & 0 & 0 & - \frac{\|\tilde{h}_2\|}{2}& 0 \\
    0 & 0 & - \frac{ \kappa }{2} & 0 & 0 & 0 & 0 \\
  \end{array}
  \right)
  \ , 
  \nonumber
  \label{eq:coeffmatrixnonorth}
\end{eqnarray}
where $\kappa = \sqrt{\|\tilde{h}_3\|^2 - |(\tilde{h}_3|h_1)|^2}$ and
$\| h \| = \sqrt{(h|h)}$. 

With the matrix $G_{ij} := (g_i|g_j)$, whose matrix-elements can be computed
using the prescription  given in section~\ref{sec:overlap}, one can write the
reduced density-matrix of the impurity-spin and region~b as
\begin{eqnarray}
  (\rho_{ab})_{ij} &=  (\mathrm{tr}_x \ \rho_{axb})_{ij} \nonumber \\ 
  &= \sum_{n,m} C_{in}C_{jm} (g_n|g_m) \ , 
\end{eqnarray}
or in matrix form:
\begin{eqnarray}
  \rho_{ab}&= C G C^{\mathrm{T}} \ .
\end{eqnarray}

\subsection{An odd number of sites to be traced out}
\begin{enumerate}
  \item For the impurity-spin we choose the $S_z$-basis and  denoted the two
    states by $f_1=\, \uparrow $ and $f_2=\, \downarrow$ .

  \item For the region to be traced out we denote the basis by ${g_i}$. 
    The states are chosen to be:
    \begin{eqnarray}
      g_1=& \underbrace{  \relbar \ \dots \ \relbar \downarrow}_{2R+1\ \mathrm{sites}} \ , 
      \ \  
      g_2= \underbrace{  \relbar \ \dots \ \relbar \downarrow}_{2R+1\ \mathrm{sites}} \ , \nonumber \\
      g_3=& \sum_{n=1}^{R} a_n \  \underbrace{\uparrow\ \overbrace{\ \relbar
      \phantom{\uparrow} \dots \phantom{\uparrow} \relbar \ }^{n - 1\ \mathrm{dimers}}\
      \uparrow \ \relbar \ \dots \ \relbar \uparrow }_{2R+1\ \mathrm{ sites}} \ , \nonumber \\
      g_4=& \sum_{n=1}^{R} a_n \  \underbrace{\downarrow\ \overbrace{\ \relbar
      \phantom{\uparrow} \dots \phantom{\uparrow} \relbar \ }^{n-1\ \mathrm{dimers}}\
      \uparrow \ \relbar \ \dots \ \relbar \uparrow }_{2R+1\ \mathrm{ sites}} \ , \nonumber \\
      g_5=& \sum_{n=1}^{R} a_n \  \underbrace{\uparrow\ \overbrace{\ \relbar
      \phantom{\uparrow} \dots \phantom{\uparrow} \relbar \ }^{n - 1\ \mathrm{dimers}}\
      \uparrow \ \relbar \ \dots \ \relbar \downarrow }_{2R+1\ \mathrm{ sites}} \ , \nonumber \\
      g_6=& \sum_{n=1}^{R} a_n \  \underbrace{\downarrow\ \overbrace{\ \relbar
      \phantom{\uparrow} \dots \phantom{\uparrow} \relbar \ }^{n-1\ \mathrm{dimers}}\
      \uparrow \ \relbar \ \dots \ \relbar \downarrow }_{2R+1\ \mathrm{ sites}} \ , \nonumber \\
      g_7=& \underbrace{  \uparrow \relbar \ \dots \ \relbar }_{2R+1\ \mathrm{sites}} \ , 
      \ \  
      g_8= \underbrace{ \downarrow \relbar \ \dots \ \relbar }_{2R+1\ \mathrm{sites}} \ , \nonumber 
    \end{eqnarray}
    where the symbol '$\relbar$' is used for valence-bonds between neighbouring
    sites and $R$ is the highest possible number of dimers that can could be
    formed in region~x.

  \item For region~b we define the states in the following way:
    \begin{eqnarray}
      \tilde{h}_1=& \underbrace{  \uparrow \relbar \ \dots \ \relbar }_{L-2R-2\ \mathrm{sites}} \ , 
      \ \  
      \tilde{h}_2= \underbrace{ \downarrow \relbar \ \dots \ \relbar }_{L-2R-2\ \mathrm{sites}} \ , \nonumber  \\
      \tilde{h}_3=& \sum_{n=R+1}^{N-1} a_n \  \underbrace{ \overbrace{\ \relbar
      \phantom{\uparrow} \dots \phantom{\uparrow} \relbar \ }^{n-R-1\ \mathrm{dimers}}\
      \uparrow \ \relbar \ \dots \ \relbar}_{L-2R-2\ \mathrm{ sites}}  \ . \nonumber 
    \end{eqnarray} 
    We orthonormalize and build a product-basis for the impurity-spin and region~b just
    as it is done in the case of an even number of sites that have to be traced out.

\end{enumerate}

With  $G_{ij} := (g_i|g_j)$ and
\begin{eqnarray}
  C =
  \left(
  \begin{array}{cccccccc}
    - \frac{a_0}{\sqrt{2}} & 0 & 0 & 0 & 0 & -\frac{1}{2} & 0 &\frac{(\tilde{h}_1|\tilde{h}_3)}{\sqrt{2}} \\
    0 &  \frac{a_0}{\sqrt{2}}  & 0 & \frac{1}{2} & 0 & 0 & 0 & 0 \\
    0 & 0 & 0 & 0 & 0 & 0 & 0 & \frac{\kappa}{\sqrt{2}} \\
    0 & 0 & 0 & 0 & \frac{1}{2} & 0 & -\frac{(\tilde{h}_1|\tilde{h}_3)}{\sqrt{2}}  & 0 \\
    0 & 0 & -\frac{1}{2} & 0 & 0 & 0 & 0 & 0 \\ 
    0 & 0 & 0 & 0 & 0 & 0 & - \frac{\kappa}{\sqrt{2}} & 0 \\
  \end{array}
  \right)
  \ ,
  \nonumber
\end{eqnarray}
where $\kappa = \sqrt{(\tilde{h}_3|\tilde{h}_3) - |(\tilde{h}_3|h_1)|^2}$, we
can again write the reduced density-matrix as
\begin{eqnarray}
  \rho_{ab}&= C G C^{\mathrm{T}} \ .
\end{eqnarray}
For a chain with an even total number of sites we followed an analogous
procedure. In order to save space we have not included it here.

\subsection{The Negativity}
With the expression for the reduced density-matrix of the impurity-spin
and region~b, $\rho_{ab}$, at hand it is now straight forward to calculate
the negativity according to its definition \cite{vidal_computable_2002}
\begin{eqnarray}
  \mathcal{N} =\frac{ \sum | \lambda_i | - 1}{2}  \ , 
\end{eqnarray}
where the $\lambda_i$ are the eigenvalues of the partial transpose of $\rho_{ab}$.

\subsection{Precision of the Variational Approach}
The variational states that form our starting point should yield almost
exact results close to the MG-point ($J_2=J/2$) where a chain with an
even number of spins is fully dimerized. However, the states we include
in our variational calculation  (see section~\ref{sec:spaces}) form only
a subset of the states in which the bulk of the chain is dimerized. At
the MG-point the contribution arising from these neglected states is
negligible and the variational approach is very good. However, as $J_2$
is decreased toward the critical point $J_2^c$ we expect the
contribution from the neglected states to grow in importance and at
$J_2^c$, where the ground-state is no longer dimerized, we expect our
variational approach to fail.  Nevertheless, for intermediate $J_2$,
$J_2^c<J_2\leq J/2$ we still expect the variational approach to yield
qualitatively good results.
\begin{figure}[h]
  \begin{center}
    \includegraphics[width=0.8\textwidth]{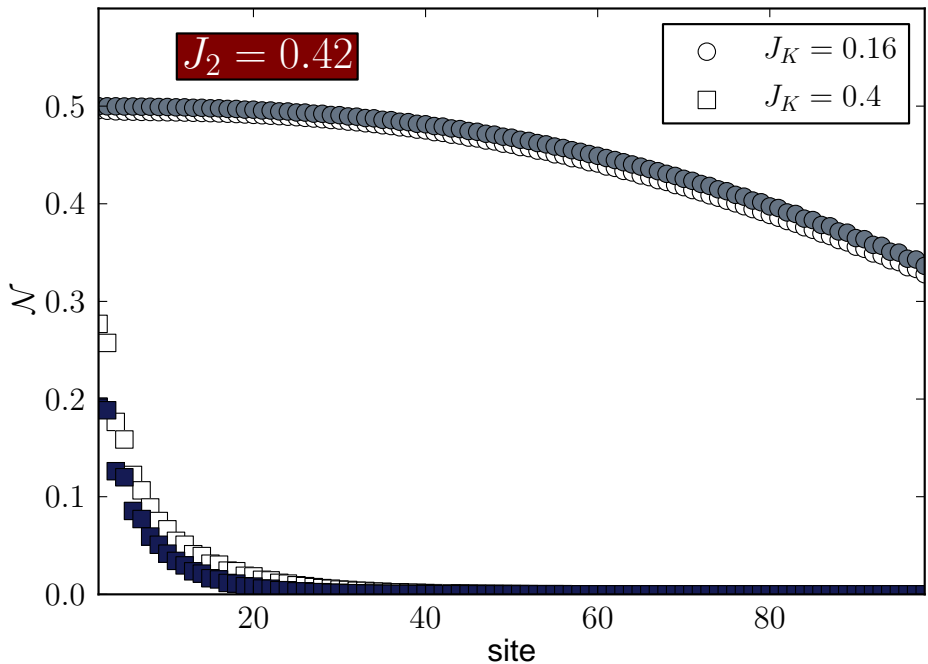}
  \end{center}
  \caption{Variational results for the negativity (filled symbols)
  together with DMRG-data from reference~\cite{bayat_negativity_2010} (empty
  symbols) for two values of the impurity-coupling $J_K$. The
  calculations were done for a chain of length $L=250$, with 
  frustration $J_2 = 0.42$ and dimerization $\delta = 0 $.}
  \label{fig:negdmrg}
\end{figure}

It is possible to calculate the negativity, $\mathcal{N}$, using DMRG
techniques, as was shown by Bayat \etal~\cite{bayat_negativity_2010}.  In
figure~\ref{fig:negdmrg} we compare our variational results for $\mathcal{N}$
to DMRG results from reference~\cite{bayat_negativity_2010} at $J_2=0.42J$ for
2 different values of $J_K$: $J_K=0.16$ and $J_K=0.4$ (Here $\delta=0$). The
definition of $\mathcal{N}$ used in reference~\cite{bayat_negativity_2010}
differs from the one used by us by a factor of two. To account for this
difference we multiplied their data by two. For both values of $J_K$ we observe
a very good agreement between the DMRG and variational results.

\section{$\mathcal{N}$ Compared to $S_{\mathrm{imp}}$ at the MG-Point}
\label{sec:negandsimp}
\begin{figure}[h]
  \begin{center}
   \includegraphics[width=0.8\textwidth]{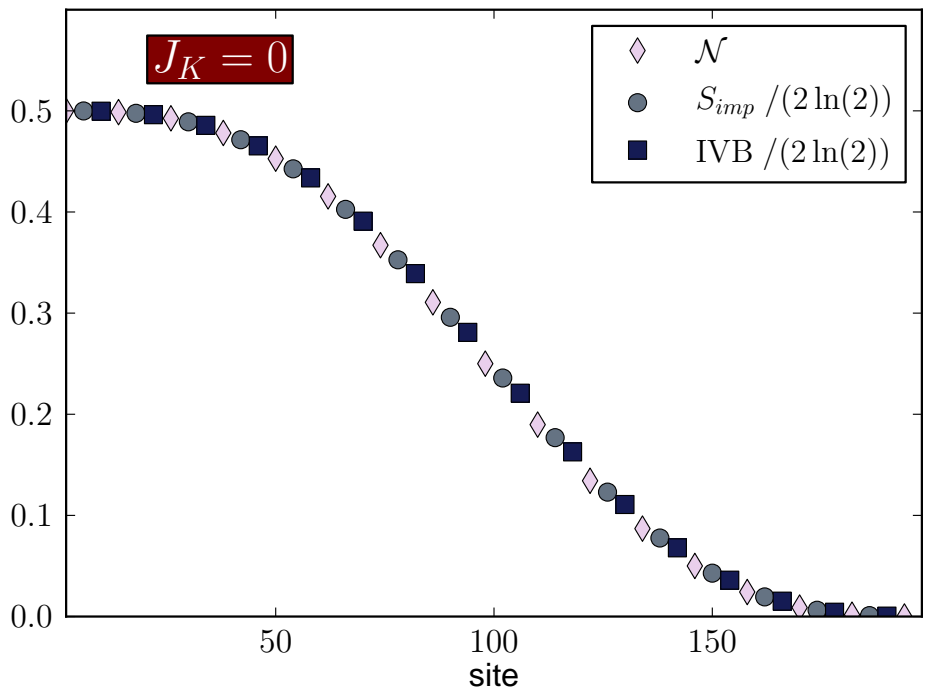}
  \end{center}
  \caption{Variational results for the negativity $\mathcal{N}$ and the
  impurity entanglement $S_{imp}$ scaled by $1/ (2 \mathrm{ln} (2))$ at
  impurity-coupling $J_K=0$ and $\delta=0$. Shown is also the impurity
  entanglement arising from an impurity valence bond (IVB) scaled by $1/
  (2 \mathrm{ln} (2))$.  The calculations were done for a chain of
  length $L=200$ at the MG-point. For clarity only a fraction of all
  data points are shown.}
  \label{fig:simpnegjk0}
\end{figure}

Having established the validity of the variational approach we now turn
to a comparison of the two measures for the impurity entanglement
$\mathcal{N}$~\cite{bayat_negativity_2010} and
$S_{\mathrm{imp}}$~\cite{sorensen_impurity_2007}.  In
figure~\ref{fig:simpnegjk0} we show results for both $\mathcal{N}$ and
$S_{\mathrm{imp}}$ at the MG-point for a system with $L=200$ sites and
an impurity-coupling $J_K=0$. Shown is also the impurity entanglement
arising from a single {\it impurity valence bond}
(IVB)~\cite{sorensen_impurity_2007} between the impurity-spin and the
bulk of the chain. 
For $J_K=0$ and $L$ even, the idea of an impurity valence bond is easy
to understand. The uncoupled impurity must form a singlet with the
unpaired $S=1/2$ in the bulk of the chain. If the entire system
(impurity and bulk) is divided in 2 parts $A$ and $B$ this impurity
valence bond will contribute either $\ln(2)$ or $0$ depending on whether
the unpaired spin is in part $B$ or $A$. With
the wave-function for the unpaired spin $\varphi_{\mathrm{sol}}(x)$ and
$p=\int_A|\varphi_{\mathrm{sol}}(x)|^2dx$, the impurity entanglement
is then simply~\cite{sorensen_impurity_2007,sorensen_quantum_2007}:
\begin{equation}
  S_{\mathrm{imp}}^{\mathrm{IVB}} = (1-p)\ln(2).\label{eq:ivb}
\end{equation}
To a first approximation $\varphi_{\mathrm{sol}}(x)$ can be taken to be the
wave-function of a free particle in a box~\cite{sorensen_quantum_2007}:
\begin{equation}  
  \varphi_{\mathrm{sol}}(i) = \sqrt{\frac{2}{L}}\mathrm{sin}\left(\frac{\pi i }{L}\right) \ .  
  \label{solwave}
\end{equation}
It is also possible to obtain  more precise estimates of
$\varphi_{\mathrm{sol}}$~\cite{sorensen_quantum_2007}.  Apart from an overall
scaling factor of $2\ln(2)$ the two measures are in the case shown in figure
\ref{fig:simpnegjk0} essentially identical.
\begin{figure}[h]
  \begin{center}
    \subfigure[$L=200$]{\includegraphics[width=0.48\textwidth]{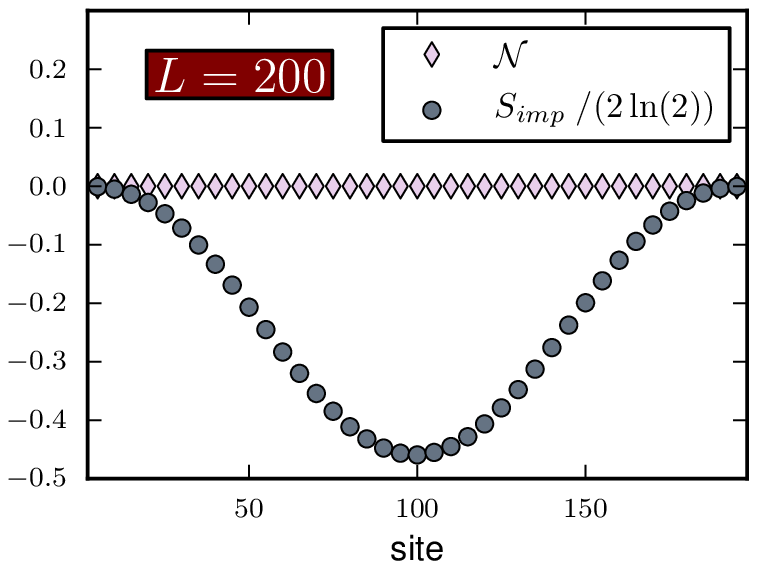}}
    \subfigure[$L=199$]{\includegraphics[width=0.48\textwidth]{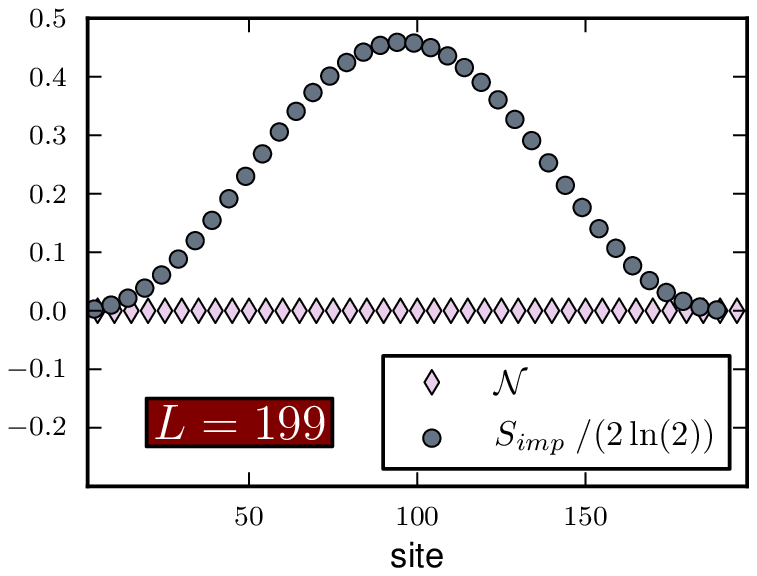}}
  \end{center}
  \caption{ Variational results for the negativity $\mathcal{N}$ and the
  impurity entanglement $S_{imp}$ scaled by $1/ (2 \mathrm{ln} (2))$ at
  impurity-coupling $J_K=1$.  The calculations were done at the MG-point for a
  chain of length $L=200$ (a) as well as $L=199$ (b).  For clarity only a
  fraction of all data points are shown.}
  \label{fig:simpnegjk1}
\end{figure}

We now compare the two measures of the impurity entanglement in the limit
$J_K=1$.  In figure~\ref{fig:simpnegjk1}(a) we show variational results for
both quantities at $J_K=1$. Since the impurity here is coupled with the same
strength as the remaining sites and since we consider a chain with an even
number of sites, the chain is in its dimerized ground-state. The dimerization
of the state implies that every spin is maximally entangled with one of its
neighbours. It follows that the impurity is only entangled with its neighbour
and not with any other spin.  While this can be inferred from the negativity
shown in figure~\ref{fig:simpnegjk1}(a) which is exactly 0 for $l>0$, this is
not the case for $S_{\mathrm{imp}}$ which in fact becomes negative (!).  The
difference in the two measures arises from the subtractive procedure used to
define $S_{\mathrm{imp}}$ where the term $S_{U}(1,l-1,L-1)$ in addition to
terms cancelling out $S_{U}(1,l,L)$ also contains a contribution  arising from
the single unpaired spin which is present because $L-1$ is odd. This
contribution from $S_{U}(1,l-1,L-1)$ can be identified as the single particle
entanglement~\cite{sorensen_quantum_2007} (SPE).  Hence, $S_{\mathrm{imp}}$
fails in this case to be a good measure of the impurity entanglement since
$S_{U}(1,l-1,L-1)$ is not a good reference state and cannot be identified with
$S({\rm no\ impurity})$.  It would be very interesting to define a measure of
the impurity entanglement based on a relative entropy but again one would need
a well defined reference state with the impurity absent.

For a system with an {\it odd} number of sites the SPE is the only contribution
to $S_{\mathrm{imp}}$ at $J_K=1$.  Here the impurity is viewed in the general
sense as arising from the boundary conditions yielding the non-zero
$S_{\mathrm{imp}}$. However, the negativity, since it is really a tri-partite
measure, is not sensitive to this and is negligible for $l>1$ as shown in
figure~\ref{fig:simpnegjk1}(b) at the MG-point.

It may be conjectured that $\mathcal{N}$ and $S_\mathrm{imp}$ agree well as
long as the IVB contribution dominates $S_\mathrm{imp}$ as is the case in
figure~\ref{fig:simpnegjk0}. We also note that the SPE vanishes at
$J_2^c$~\cite{sorensen_quantum_2007} and we expect $\mathcal{N}$ and
$S_{\mathrm{imp}}$ to exhibit the same scaling behaviour at $J_2^c$ as it
indeed has been
demonstrated~\cite{sorensen_impurity_2007,bayat_negativity_2010}.  For
$J_2>J_2^c$, $\mathcal{N}$ is the more faithful measure of the impurity
entanglement and in the following we mainly use this measure.

\section{The Negativity and the SPE for general $J_K$ at the MG-Point}
\label{sec:negandjk}
\begin{figure}[h]
  \begin{center}
     \subfigure{\includegraphics[width=0.48\textwidth]{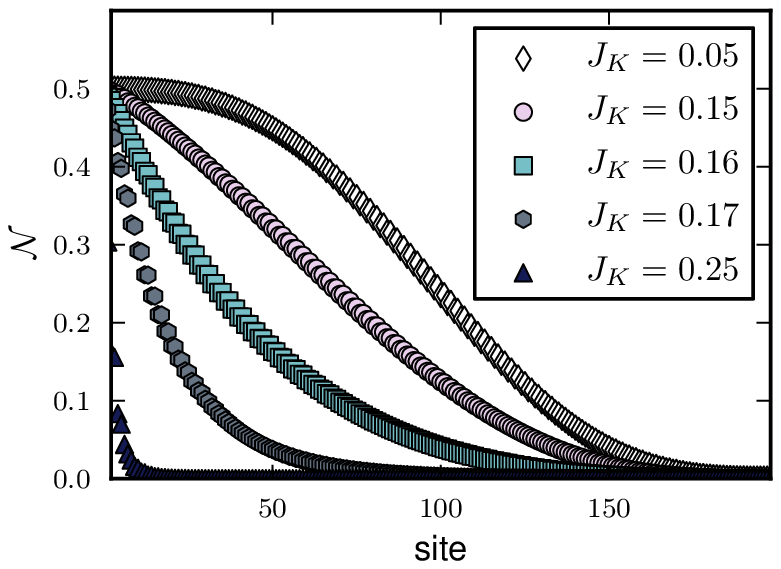}}
     \subfigure{\includegraphics[width=0.48\textwidth]{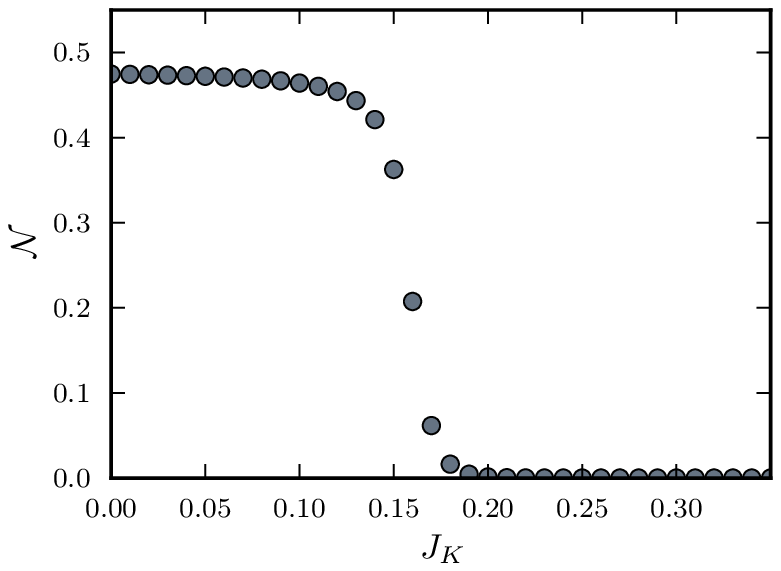}}
  \end{center}
  \caption{(a) Variational results for the negativity for different
  $J_K$ with $L=200$. (b) Variational results for the negativity for
  fixed size, $l=40$, of the subsystem x as a function of $J_K$ with
  $L=200$. All results are obtained at the MG-point with $\delta=0$. }
  \label{fig:crossoverIVB}
\end{figure}
As we saw in the previous section, the impurity entanglement for an {\it
even} length system with $J_K=0$ can be seen as arising from an impurity
valence bond (IVB). At the MG-point for $J_K=0$ this was calculated
within the variational approach in
reference~\cite{sorensen_quantum_2007}.  Here we focus on the negativity
$\mathcal{N}$ and the complete cross-over as $J_K$ is varied between $0$
and $1$. Variational results of $\mathcal{N}$ at different $J_K$ for a
system of length $L=200$ (including the impurity site) are shown in
figure~\ref{fig:crossoverIVB}(a). At $J_K=0$ the impurity entanglement
is long-range, extending throughout the chain, but as $J_K$ is turned on
it quickly becomes suppressed and for $J_K$ larger than $\sim0.3$ it has
all but disappeared reflecting the fact that the impurity valence bond
now preferentially terminates close to the impurity site. For these
non-zero values of $J_K$ the appropriate $\varphi_{sol}$ to use would
then no longer be the wave-function of a free particle, as in
equation~(\ref{solwave}), but instead a wave-function describing a
localized state bound to the impurity site.  The presence of such a
localized state should be reflected in an exponentially decaying
impurity entanglement away from the impurity site.  For $J_K > 0.15$ we
have verified that the negativity $\mathcal{N}$ shown in
figure~\ref{fig:crossoverIVB}(a) indeed has a tail decaying
exponentially with the site index.  The length scale,
$\xi_{\mathrm{loc}}$, associated with this exponential decay is
therefore clearly distinct from the bulk spin-spin correlation length in
the system which here is effectively zero.  The sharpness of the
cross-over is illustrated in figure~\ref{fig:crossoverIVB}(b) where for
a chain of $200$ spins the negativity with the region x containing $40$
spins is plotted versus $J_K$.  As can be seen, $\mathcal{N}$
transitions from $\sim 0.5$ to $0$ very close to $J_K=0.15$.  For
$J_K\leq 0.15$ we have not been able to identify any exponentially
decaying part in $\mathcal{N}$ as obtained within the variational
approach even for systems substantially longer than $200$ and it appears
that a finite $J_K^c\sim 0.15$ is needed to induce an exponential decay
in $\mathcal{N}$. This would imply that the presence of a non-zero $J_K$
cannot be modelled as a simple one-dimensional potential well, since
this would always have a bound-state independent of the depth of the
potential well.  Likely, an enlarged variational space for the
calculation will change the value for $J_K^c$ and could possibly drive it all
the way to zero. However, Bayat \etal~\cite{bayat_negativity_2010}, using DMRG, do not
find an exponential decay for $J_2=0.46$ and $J_K=0.16$ consistent with a non-zero
$J_K^c$. Due to the
complexity of the calculation of $\mathcal{N}$ in larger sub-spaces we have, however, been
unable to complete such calculations. 
\begin{figure}[h]
  \begin{center}
     \includegraphics[width=0.8\textwidth]{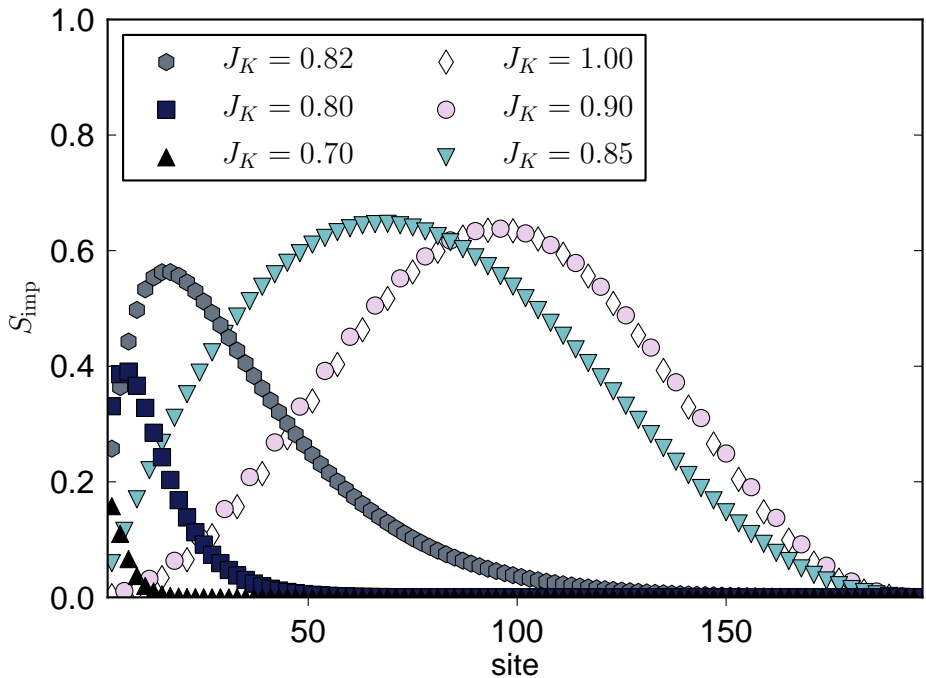}
  \end{center}
  \caption{Variational results for the $S_{\mathrm{imp}}$ for different
  $J_K$ with $L=199$. All results are obtained at the MG-point with
  $\delta=0$. For clarity only a fraction of all data points is shown.}
  \label{fig:crossoverSPE}
\end{figure}

In a similar manner we can study the single particle entanglement (SPE) for
general $J_K$ within the variational approach.  Our results are shown in
figure~\ref{fig:crossoverSPE}.  The single particle entanglement arises from
the presence of a single unpaired spin for odd $L$ which is always present in
the ground-state for odd $L$. The unpaired spin moves freely throughout the
system when $J_K=1$ but as $J_K$ is decreased away from 1 towards zero it
rapidly becomes localized on the impurity site thereby quenching the impurity
entanglement.  As can be seen in figure~\ref{fig:crossoverSPE} the impurity
entanglement rapidly transitions around $J_K\sim0.8-0.85$. While the
variational approach agrees closely with DMRG data away from this transition
region, some quantitative differences appear in this region due to the
limitation of the Hilbert space used in the variational approach.  The
variational and DMRG results are however qualitatively the same and we have for
clarity not included DMRG results in figure~\ref{fig:crossoverSPE}.  We have
verified that for $J_K \leq 0.85$ the data for $S_{\mathrm{imp}}$ shown in
figure~\ref{fig:crossoverSPE} develop an exponentially decaying part with an
associated length scale, $\xi_{\mathrm{loc}}$, different from the bulk
spin-spin correlation length.  Also here, it appears that $\xi_{\mathrm{loc}}$
diverges before $J_K$ reaches 1 but this effect could possibly be due to
finite-size effects in the DMRG calculations or limitations in the sub-space
used in the variational calculations.

\section{Impurity Entanglement with $\delta\neq 0$}
\label{sec:impentdeluneq0}
\begin{figure}[h]
  \begin{center}
   \subfigure{\includegraphics[width=0.48\textwidth]{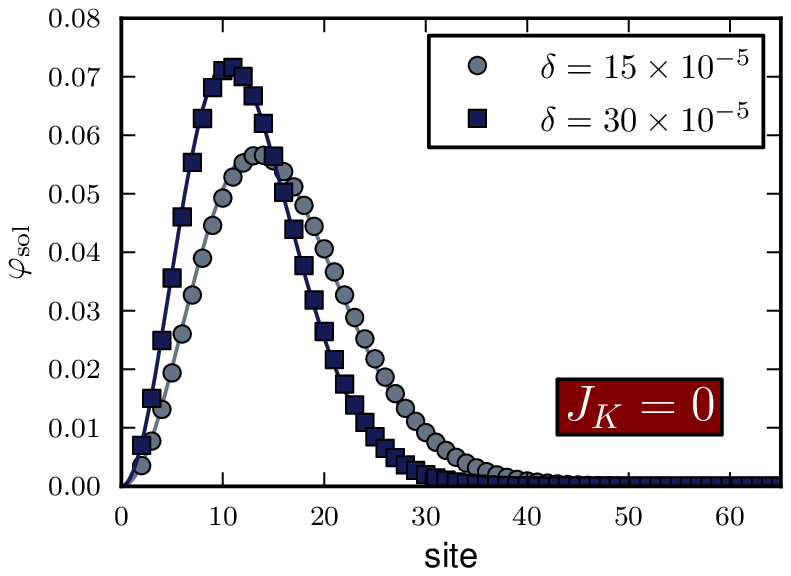}}
   \subfigure{\includegraphics[width=0.48\textwidth]{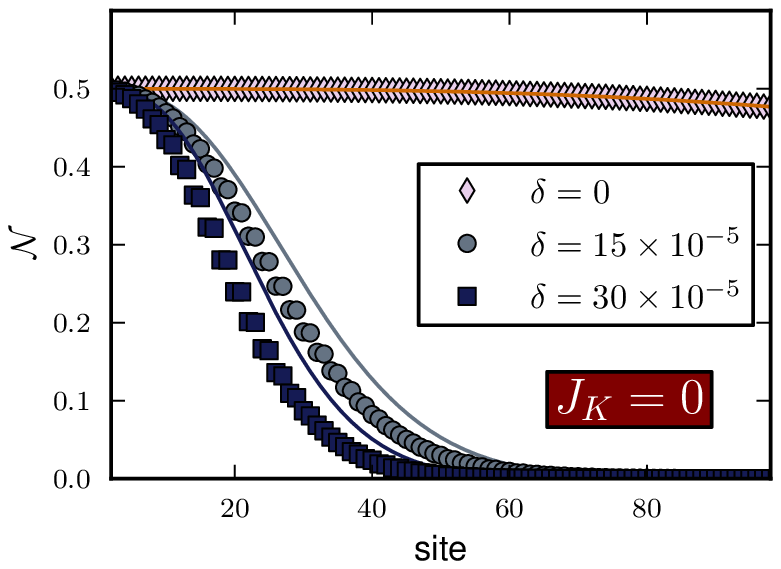}}
  \end{center}
  \caption{(a) Variational results for $\varphi_{\mathrm{sol}}$ (markers)
  compared to $\mathrm{Ai}(i / \xi + z )$ (lines) (see main text).  (b)
  Variational results for the negativity at impurity-coupling $J_K=0$ for three
  values of the dimerization $\delta$ plotted along with the scaled
  IVB~entanglement (lines).  The calculations were done for a chain of length
  $L=500$, with $J_2=J/2$.}
  \label{fig:negtrans}
\end{figure}
The variational approach that we employ here is straight forward to use
for any $J_2, \delta$ and $J_K$ and we now discuss our results for the
impurity entanglement for non-zero explicit dimerization $\delta\neq 0$.
As we stressed in the introduction the impurity entanglement can be
long-range in these systems due to the near degeneracy of the two
singlet ground-states of the chain with periodic boundary conditions.
With a non-zero $\delta$ this degeneracy is lifted and we expect the
impurity entanglement to decrease dramatically.  Essentially, this is
because when $\delta$ is increased it is more costly for the impurity to
be entangled with far away parts of the systems: If the impurity is in a
singlet with a spin far away, there are more dimers on weak bonds than
if the impurity bonds with a site that is close to it.  If we consider
$L$ even, then, in order to have sizeable entanglement, the bulk of the
chain must have a site in a valence-bond with the impurity-spin and this
spin is then bound to one end of the chain~\cite{sorensen_soliton_1998}.

For a chain with an even number of spins we expect to be able to describe
$\mathcal{N}$ in terms of an IVB picture for which an estimate of
$\varphi_{\mathrm{sol}}$ is needed for non-zero $\delta$.  Such an estimate has
been derived by Uhrig \etal (reference~\cite{uhrig_unified_1999}).  They showed
that to good approximation $\varphi_{\mathrm{sol}}(i) \propto \mathrm{Ai}(i /
\xi + z) $, where $\mathrm{Ai(x)}$ is the Airy-function and $z_1=-2.3381$ is
its biggest root. Here, the size of the area the impurity-spin binds to is
roughly given by $\xi = (3 m  \delta/ 2)^{-1/ 3}$, where $m \approx 1/ (1 + 7 /
\sqrt{65} )$. Our variational approach directly yields the lowest state vector
which is simply $\varphi_{\mathrm{sol}}$.  In figure~\ref{fig:negtrans}(a) we
compare this directly to the $\mathrm{Ai}(i / \xi + z ) $ where it can be seen
that the agreement is good. It is noteworthy that the maximum in
$\varphi_{\mathrm{sol}}$ is quite distant from the end of the chain in
agreement with previous results.

We can use this result to estimate $\mathcal{N}$ using equation~(\ref{eq:ivb}).
In figure~\ref{fig:negtrans}(b) we see typical graphs of the behaviour of the
negativity as $\delta$ is increased with $J_K=0$ and $L=500$.  The range over
which the impurity-site is entangled is decreased upon increasing the
dimerization. Figure~\ref{fig:negtrans}(b) also includes the impurity valence
bond entropy (IVB) obtained from equation~(\ref{eq:ivb}) using the Airy
functions for $\varphi_{\mathrm{sol}}$.  For easier comparison with the
negativity we scaled by a factor of $1/ 2 \, \ln2$. For $\delta = 0$ we used
equation~(\ref{solwave}) for $\varphi_{\mathrm{sol}}$.  A reasonable agreement
between $\mathcal{N}$ and the IVB contribution (solid lines) is observed.
Overall, as one would expect, the impurity entanglement is now dramatically
suppressed. Even for $\delta=15\times 10^{-5}$ the impurity entanglement is
effectively zero beyond 60 sites. Following the above discussion we expect the
extent over which the impurity entanglement is non-zero to diverge as
$\delta^{-1/3}$ as dictated by $\xi$.

\section{Conclusion}
We have presented detailed discussion of how variational calculations can be
used to calculate the negativity. This approach is quite generally applicable
and should be very precise close to the MG-point. A complete characterization
of the impurity entanglement for any $J_K$ can then be obtained. For $J_K=0$
($L$ even) the impurity entanglement is long-range and extends throughout the
system while for larger $J_K>0.15$ it decays exponentially with a length scale
clearly different from the correlation length in the system.  For $L$ odd the
SPE present at $J_K=1$ is rapidly destroyed once $J_K$ is decreased below
$J_K=0.8-0.85$. In both cases it would be valuable to perform higher precision
calculations in order to determine whether or not a critical $J_K$ or $(1-J_K)$
is needed to destroy the impurity entanglement. When a non-zero dimerization is
introduced and the ground-state degeneracy is lifted, the impurity entanglement
rapidly disappears. It would be interesting to investigate impurity
entanglement in other gapped systems with degenerate singlet ground-states.

\ack
We are grateful to I. Affleck and N. Laflorencie for interesting
discussions and prior collaborations that initiated this work.  We also
acknowledge fruitful discussions with A. Bayat, S. Bose and P. Sodano
and thank them for making their DMRG data for the negativity available
to us.
This research was supported by NSERC and CFI and 
was made possible by the facilities of the Shared Hierarchical
Academic Research Computing Network (SHARCNET:www.sharcnet.ca).

\vspace{15mm}
\section*{References}
\end{document}